# Instantaneous and lagged measurements of linear and nonlinear dependence between groups of multivariate time series: frequency decomposition

Roberto D. Pascual-Marqui

The KEY Institute for Brain-Mind Research
University Hospital of Psychiatry
Lenggstr. 31, CH-8032 Zurich, Switzerland
pascualm at key.uzh.ch
www.keyinst.uzh.ch/loreta

## 1. Abstract

Measures of linear dependence (coherence) and nonlinear dependence (phase synchronization) between any number of multivariate time series are defined. The measures are expressed as the sum of lagged dependence and instantaneous dependence. The measures are non-negative, and take the value zero only when there is independence of the pertinent type. These measures are defined in the frequency domain and are applicable to stationary and non-stationary time series. These new results extend and refine significantly those presented in a previous technical report (Pascual-Marqui 2007, arXiv:0706.1776 [stat.ME], http://arxiv.org/abs/0706.1776 ), and have been largely motivated by the seminal paper on linear feedback by Geweke (1982 JASA 77:304-313). One important field of application is neurophysiology, where the time series consist of electric neuronal activity at several brain locations. Coherence and phase synchronization are interpreted as "connectivity" between locations. However, any measure of dependence is highly contaminated with an instantaneous, non-physiological contribution due to volume conduction and low spatial resolution. The new techniques remove this confounding factor considerably. Moreover, the measures of dependence can be applied to any number of brain areas jointly, i.e. distributed cortical networks, whose activity can be estimated with eLORETA (Pascual-Marqui 2007, arXiv:0710.3341 [math-ph], http://arxiv.org/abs/0710.3341 ).

## 2. Introduction

This study extends and refines significantly the results presented in a previous technical report (Pascual-Marqui 2007a). Some results from that previous paper will be repeated here for the sake of completeness.

### 2.1. The discrete Fourier transform for multivariate time series

The terms "multivariate time series", "multiple time series", and "vector time series" have identical meaning in this paper.

For general notation and definitions, see e.g. Brillinger (1981) for stationary multivariate time series analysis, and see e.g. Mardia et al (1979) for general multivariate statistics.

Let $\mathbf{X}_{jt} \in \mathbb{R}^{p \times 1}$ and $\mathbf{Y}_{jt} \in \mathbb{R}^{q \times 1}$ denote two stationary multivariate time series, for discrete time $t = 0 \ldots N_T - 1$, with $j = 1 \ldots N_R$ denoting the $j$-th time segment. The discrete Fourier transforms are denoted as $\mathbf{X}_{j\omega} \in \mathbb{C}^{p \times 1}$ and $\mathbf{Y}_{j\omega} \in \mathbb{C}^{q \times 1}$, and defined as:

**Eq. 1**
$$\mathbf{X}_{j\omega} = \sum_{t=0}^{N_T - 1} \mathbf{X}_{jt} e^{-2\pi i \omega t / N_T}$$

**Eq. 2**
$$\mathbf{Y}_{j\omega} = \sum_{t=0}^{N_T - 1} \mathbf{Y}_{jt} e^{-2\pi i \omega t / N_T}$$

for discrete frequencies $\omega = 0 \ldots N_T - 1$, and where $i = \sqrt{-1}$.

It will be assumed throughout that $\mathbf{X}_\omega$ and $\mathbf{Y}_\omega$ each have zero mean.

### 2.2. Classical cross-spectra

Let:

**Eq. 3**
$$\mathbf{S}_{XX\omega} = \frac{1}{N_R} \sum_{j=1}^{N_R} \mathbf{X}_{j\omega} \mathbf{X}_{j\omega}^*$$

**Eq. 4**
$$\mathbf{S}_{YY\omega} = \frac{1}{N_R} \sum_{j=1}^{N_R} \mathbf{Y}_{j\omega} \mathbf{Y}_{j\omega}^*$$

**Eq. 5**
$$\mathbf{S}_{XY\omega} = \frac{1}{N_R} \sum_{j=1}^{N_R} \mathbf{X}_{j\omega} \mathbf{Y}_{j\omega}^*$$

**Eq. 6**
$$\mathbf{S}_{YX\omega} = \mathbf{S}_{XY\omega}^* = \frac{1}{N_R} \sum_{j=1}^{N_R} \mathbf{Y}_{j\omega} \mathbf{X}_{j\omega}^*$$

denote complex valued covariance matrices, where the superscript "*" denotes vector/matrix transposition and complex conjugation. Note that $\mathbf{S}_{XX\omega}$ and $\mathbf{S}_{YY\omega}$ are Hermitian matrices, satisfying $\mathbf{S} = \mathbf{S}^*$. When multiplied by the factor $(2\pi N_T)^{-1}$, these matrices correspond to the classical cross-spectral density matrices.

### 2.3. Phase-information cross-spectra

The discrete Fourier transforms in Eq. 1 and Eq. 2 contain both phase and amplitude information, which carries over to the covariance matrices in Eq. 3, Eq. 4, Eq. 5, and Eq. 6. This means that for the analysis of phase information, the amplitudes must be factored out by an appropriate normalization method. This is achieved by using the following definition for the normalized complex-valued discrete Fourier transform vector:

**Eq. 7**
$$\breve{\mathbf{X}}_{j\omega} = \left(\mathbf{X}_{j\omega}^* \mathbf{X}_{j\omega}\right)^{-1/2} \mathbf{X}_{j\omega}$$

and:

Eq. 8 $\quad \breve{\mathbf{Y}}_{j\omega} = \left(\mathbf{Y}_{j\omega}^{*}\mathbf{Y}_{j\omega}\right)^{-1/2} \mathbf{Y}_{j\omega}$

Note that this normalization operation, although deceivingly simple, is a highly nonlinear transformation.

The corresponding covariance matrices containing phase information (without amplitude information) are:

Eq. 9 $\quad \mathbf{S}_{\breve{\mathbf{X}}\breve{\mathbf{X}}\omega} = \frac{1}{N_R}\sum_{j=1}^{N_R}\breve{\mathbf{X}}_{j\omega}\breve{\mathbf{X}}_{j\omega}^{*}$

Eq. 10 $\quad \mathbf{S}_{\breve{\mathbf{Y}}\breve{\mathbf{Y}}\omega} = \frac{1}{N_R}\sum_{j=1}^{N_R}\breve{\mathbf{Y}}_{j\omega}\breve{\mathbf{Y}}_{j\omega}^{*}$

Eq. 11 $\quad \mathbf{S}_{\breve{\mathbf{X}}\breve{\mathbf{Y}}\omega} = \frac{1}{N_R}\sum_{j=1}^{N_R}\breve{\mathbf{X}}_{j\omega}\breve{\mathbf{Y}}_{j\omega}^{*}$

Eq. 12 $\quad \mathbf{S}_{\breve{\mathbf{Y}}\breve{\mathbf{X}}\omega} = \mathbf{S}_{\breve{\mathbf{X}}\breve{\mathbf{Y}}\omega}^{*} = \frac{1}{N_R}\sum_{j=1}^{N_R}\breve{\mathbf{Y}}_{j\omega}\breve{\mathbf{X}}_{j\omega}^{*}$

Note that the normalization used in Eq. 7 and Eq. 8 will be the basis for the analysis of phase synchronization between the multivariate time series **X** and **Y**.

Note that $\mathbf{S}_{\breve{\mathbf{X}}\breve{\mathbf{X}}\omega}$ and $\mathbf{S}_{\breve{\mathbf{Y}}\breve{\mathbf{Y}}\omega}$ are Hermitian matrices. When multiplied by the factor $\left(2\pi N_T\right)^{-1}$, these matrices correspond to what is defined here as the phase-information cross-spectra.

### 2.4. Instantaneous, zero-phase, zero-lag covariance

The instantaneous, zero-phase, zero-lag covariance matrix corresponding to a multivariate time series at frequency $\omega$, is simply the real part of the Hermitian covariance matrix at frequency $\omega$, i.e. $\text{Re}(\mathbf{S}_\omega)$.

To justify this, consider the multivariate time series $\mathbf{X}_{jt} \in \mathbb{R}^{p \times 1}$, for discrete time $t = 0...N_T - 1$, with $j = 1...N_R$ denoting the $j$-th time segment.

In a first step, filter the time series to leave exclusively the frequency $\omega$ component. Denote the filtered time series as $\left(\mathbf{X}_{jt}^{\omega Filtered}\right)$. Note that, by construction, the spectral density of $\left(\mathbf{X}_{jt}^{\omega Filtered}\right)$ is zero everywhere except at frequency $\omega$.

In a second step, compute the instantaneous, zero-lag, zero phase shifted, time domain, symmetric covariance matrix for the filtered time series $\left(\mathbf{X}_{jt}^{\omega Filtered}\right)$ at frequency $\omega$:

Eq. 13 $\quad \mathbf{A}_\omega = \frac{1}{N_T N_R}\sum_{j=1}^{N_R}\sum_{t=1}^{N_t}\left(\mathbf{X}_{jt}^{\omega Filtered}\right)\left(\mathbf{X}_{jt}^{\omega Filtered}\right)^T \in \mathbb{R}^{r \times r}$

Finally, by making use of Parseval's theorem for the filtered time series, the following relation holds:

**Eq. 14**    $$\mathrm{Re}(\mathbf{S}_{XX\omega}) = \frac{N_T^2}{2}\mathbf{A}_\omega$$

where $\mathrm{Re}(\mathbf{S}_{XX\omega})$ denotes the real part of $\mathbf{S}_{XX\omega}$ given by Eq. 3 above.

These arguments apply identically to the normalized time series, as in Eq. 7 to Eq. 12 above, when considering the phase-information cross-spectra. This means that the instantaneous, zero-phase, zero-lag covariance matrix corresponding to a normalized multivariate time series $\mathbf{X}$ at frequency $\omega$, is simply the real part of the phase-information Hermitian covariance matrix at frequency $\omega$, i.e. $\mathrm{Re}(\mathbf{S}_{\tilde{X}\tilde{X}\omega})$.

The section entitled "Appendix 1" gives a brief description of the problems that arise in neurophysiology, where any measure of dependence is highly contaminated with an instantaneous, non-physiological contribution due to volume conduction and low spatial resolution.

## 3. Measures of linear dependence (coherence-type) between two multivariate time series

The definitions presented here are largely motivated by the seminal paper on linear feedback by Geweke (1982).

The measure of linear dependence between time series $\mathbf{X}$ and $\mathbf{Y}$ at frequency $\omega$ is defined as:

**Eq. 15**    $$F_{X,Y}(\omega) = \ln \frac{\left|\begin{pmatrix} \mathbf{S}_{YY\omega} & \mathbf{o} \\ \mathbf{o}^T & \mathbf{S}_{XX\omega} \end{pmatrix}\right|}{\left|\begin{pmatrix} \mathbf{S}_{YY\omega} & \mathbf{S}_{YX\omega} \\ \mathbf{S}_{XY\omega} & \mathbf{S}_{XX\omega} \end{pmatrix}\right|}$$

where $|\mathbf{M}|$ denotes the determinant of $\mathbf{M}$. The matrix in the numerator of Eq. 15 is a block-diagonal matrix, with $\mathbf{o}$ denoting a matrix of zeros, which in this case is of dimension $q \times p$.

This measure of linear dependence is expressed as the sum of the lagged linear dependence $F_{X \rightleftarrows Y}(\omega)$ and instantaneous linear dependence $F_{X \cdot Y}(\omega)$:

**Eq. 16**    $$F_{X,Y}(\omega) = F_{X \rightleftarrows Y}(\omega) + F_{X \cdot Y}(\omega)$$

The measure of instantaneous linear dependence is defined as:

**Eq. 17**    $$F_{X \cdot Y}(\omega) = \ln \frac{\left|\mathrm{Re}\begin{pmatrix} \mathbf{S}_{YY\omega} & \mathbf{o} \\ \mathbf{o}^T & \mathbf{S}_{XX\omega} \end{pmatrix}\right|}{\left|\mathrm{Re}\begin{pmatrix} \mathbf{S}_{YY\omega} & \mathbf{S}_{YX\omega} \\ \mathbf{S}_{XY\omega} & \mathbf{S}_{XX\omega} \end{pmatrix}\right|}$$

where $\mathrm{Re}(\mathbf{M})$ denotes the real part of $\mathbf{M}$.

Finally, the measure of lagged linear dependence is:

Eq. 18 $$F_{X \rightleftarrows Y}(\omega) = F_{X,Y}(\omega) - F_{X \cdot Y}(\omega) = \ln \frac{\left\{ \left| \mathrm{Re} \begin{pmatrix} \mathbf{S}_{YY\omega} & \mathbf{S}_{YX\omega} \\ \mathbf{S}_{XY\omega} & \mathbf{S}_{XX\omega} \end{pmatrix} \right| \middle/ \left| \mathrm{Re} \begin{pmatrix} \mathbf{S}_{YY\omega} & \mathbf{0} \\ \mathbf{0}^T & \mathbf{S}_{XX\omega} \end{pmatrix} \right| \right\}}{\left\{ \left| \begin{pmatrix} \mathbf{S}_{YY\omega} & \mathbf{S}_{YX\omega} \\ \mathbf{S}_{XY\omega} & \mathbf{S}_{XX\omega} \end{pmatrix} \right| \middle/ \left| \begin{pmatrix} \mathbf{S}_{YY\omega} & \mathbf{0} \\ \mathbf{0}^T & \mathbf{S}_{XX\omega} \end{pmatrix} \right| \right\}}$$

All three measures are non-negative. They take the value zero only when there is independence of the pertinent type (lagged, instantaneous, or both).

Not that the measure of linear dependence $F_{X,Y}(\omega)$ in Eq. 15 can be interpreted as follows:

Eq. 19 $$\rho_{X,Y}^2(\omega) = 1 - \exp(-F_{X,Y}(\omega))$$

where $\rho_{X,Y}(\omega)$ was defined as the general coherence in Pascual-Marqui (2007a; see Eq. 7 therein):

Eq. 20 $$\rho_{X,Y}^2(\omega) = \rho_G^2 = 1 - \frac{\left| \mathbf{S}_{YY\omega} - \mathbf{S}_{YX\omega} \mathbf{S}_{XX\omega}^{-1} \mathbf{S}_{XY\omega} \right|}{\left| \mathbf{S}_{YY\omega} \right|}$$

Some relevant literature that motivated the definition of the general coherence $\rho_{X,Y}^2(\omega)$ in the previous study (Pascual-Marqui 2007a) follows. In the case of real-valued stochastic variables, Mardia et al (1979) review several "measures of correlation between vectors". In particular, Kent (1983) proposed a general measure of correlation that is closely related to the vector alienation coefficient (Hotelling 1936, Mardia et al 1979). This measure of general coherence is also equivalent to the coefficient of determination as defined by Pierce (1982). All these definitions can be straightforwardly generalized to the complex valued domain.

In order to illustrate and further motivate these measures of linear dependence, a detailed analysis for the simple case of two univariate time series is presented.

In the case that the two time series are univariate, the measure of linear dependence $F_{X,Y}(\omega)$ in Eq. 15 is:

Eq. 21 $$F_{X,Y}(\omega) = \ln \frac{s_{yy\omega} s_{xx\omega}}{s_{yy\omega} s_{xx\omega} - \left[ \mathrm{Re}(s_{yx\omega}) \right]^2 - \left[ \mathrm{Im}(s_{yx\omega}) \right]^2} = -\ln(1 - \rho^2)$$

where:

Eq. 22 $$\rho^2 = \frac{\left( \left[ \mathrm{Re}(s_{yx\omega}) \right]^2 + \left[ \mathrm{Im}(s_{yx\omega}) \right]^2 \right)}{s_{yy\omega} s_{xx\omega}}$$

In Eq. 22, $\rho$ is the ordinary squared coherence (see e.g. Equation 3 in Nolte et al 2004).

The measure of instantaneous linear dependence is:

Eq. 23 $$F_{X \cdot Y}(\omega) = \ln \frac{s_{yy\omega} s_{xx\omega}}{s_{yy\omega} s_{xx\omega} - \left[ \mathrm{Re}(s_{yx\omega}) \right]^2}$$

Note that we can define the instantaneous coherence $\rho_{X \cdot Y}(\omega)$ as:

Eq. 24 $\quad F_{X \cdot Y}(\omega) = -\ln\left(1 - \rho_{X \cdot Y}^2(\omega)\right)$

In general, this gives:

Eq. 25 $\quad \rho_{X \cdot Y}^2(\omega) = 1 - \exp\left[-F_{X \cdot Y}(\omega)\right] = 1 - \dfrac{\left|\operatorname{Re}\begin{pmatrix} S_{YY\omega} & S_{YX\omega} \\ S_{XY\omega} & S_{XX\omega} \end{pmatrix}\right|}{\left|\operatorname{Re}\begin{pmatrix} S_{YY\omega} & \mathbf{0} \\ \mathbf{0}^T & S_{XX\omega} \end{pmatrix}\right|}$

and in the case of univariate time series it simplifies to:

Eq. 26 $\quad \rho_{X \cdot Y}^2(\omega) = \dfrac{\left[\operatorname{Re}(s_{yx\omega})\right]^2}{s_{yy\omega} s_{xx\omega}}$

which, not surprisingly, is directly related to the real part of the complex valued coherency.

Finally, in the particular case of univariate time series, the measure of lagged linear dependence is:

Eq. 27 $\quad \begin{cases} F_{X \rightleftarrows Y}(\omega) = F_{X,Y}(\omega) - F_{X \cdot Y}(\omega) \\[4pt] \qquad = \ln \dfrac{s_{yy\omega} s_{xx\omega}}{s_{yy\omega} s_{xx\omega} - \left[\operatorname{Re}(s_{yx\omega})\right]^2 - \left[\operatorname{Im}(s_{yx\omega})\right]^2} - \ln \dfrac{s_{yy\omega} s_{xx\omega}}{s_{yy\omega} s_{xx\omega} - \left[\operatorname{Re}(s_{yx\omega})\right]^2} \\[4pt] \qquad = \ln\left[\dfrac{s_{yy\omega} s_{xx\omega} - \left[\operatorname{Re}(s_{yx\omega})\right]^2}{s_{yy\omega} s_{xx\omega} - \left[\operatorname{Re}(s_{yx\omega})\right]^2 - \left[\operatorname{Im}(s_{yx\omega})\right]^2}\right] \\[4pt] \qquad = -\ln\left(1 - \rho_{X \rightleftarrows Y}^2(\omega)\right) \end{cases}$

with:

Eq. 28 $\quad \rho_{X \rightleftarrows Y}^2(\omega) = \dfrac{\left[\operatorname{Im}(s_{yx\omega})\right]^2}{s_{yy\omega} s_{xx\omega} - \left[\operatorname{Re}(s_{yx\omega})\right]^2}$

In Eq. 28, for the particular case of univariate time series, $\rho_{X \rightleftarrows Y}^2(\omega)$ is equal to the "zero-lag removed general coherence" $\rho_{GL}$ defined in Pascual-Marqui (2007a).

In our previous related study (Pascual-Marqui 2007a), the general definition given there for the "zero-lag removed coherence" (see Eq. 22 therein) was:

Eq. 29 $\quad \rho_{GL}^2 = 1 - \dfrac{\left|\begin{pmatrix} S_{YY\omega} & S_{YX\omega} \\ S_{XY\omega} & S_{XX\omega} \end{pmatrix}\right|}{\left|\operatorname{Re}\begin{pmatrix} S_{YY\omega} & S_{YX\omega} \\ S_{XY\omega} & S_{XX\omega} \end{pmatrix}\right|}$

The new definition given here for the lagged coherence follows from the relation:

Eq. 30 $\quad F_{X \rightleftarrows Y}(\omega) = -\ln\left(1 - \rho_{X \rightleftarrows Y}^2(\omega)\right)$

which gives:

Eq. 31
$$\rho^2_{X \rightleftarrows Y}(\omega) = 1 - \exp\left[-F_{X \rightleftarrows Y}(\omega)\right] = 1 - \frac{\left\{\left|\begin{pmatrix} S_{YY\omega} & S_{YX\omega} \\ S_{XY\omega} & S_{XX\omega} \end{pmatrix}\right| \middle/ \left|\begin{pmatrix} S_{YY\omega} & 0 \\ 0^T & S_{XX\omega} \end{pmatrix}\right|\right\}}{\left\{\left|\text{Re}\begin{pmatrix} S_{YY\omega} & S_{YX\omega} \\ S_{XY\omega} & S_{XX\omega} \end{pmatrix}\right| \middle/ \left|\text{Re}\begin{pmatrix} S_{YY\omega} & 0 \\ 0^T & S_{XX\omega} \end{pmatrix}\right|\right\}}$$

Both definitions (Eq. 29 and Eq. 31) are identical for the case of two univariate time series. However, they are different for the multivariate case. Whereas the old definition in Eq. 29 lumps together all variables from **X** and **Y**, the new definition given here in Eq. 31 conserves the multivariate structure of the two multivariate time series. The improvement of the new lagged coherence in Eq. 31 is that it measures the lagged linear dependence between the two multivariate time series without being affected by the covariance structure within each multivariate time series. The shortcoming of the old definition from our previous study (Pascual-Marqui 2007a), shown in Eq. 29, is that it is contaminated by the dependence structures of the univariate time series within **X** and within **Y**.

Another point worth stressing is the asymmetry in the results for the instantaneous coherence $\rho^2_{X \cdot Y}(\omega)$ (Eq. 26) and the lagged coherence $\rho^2_{X \rightleftarrows Y}(\omega)$ (Eq. 28). While the instantaneous coherence is the real part of the complex valued coherency, the lagged coherence _is not_ the imaginary part of the complex valued coherency. Ideally, the lagged coherence is a measure that is not affected by instantaneous dependence, whereas the imaginary part of the complex valued coherency (Nolte et al 2004) is more affected by instantaneous dependence (Pascual-Marqui 2007a). This makes the lagged coherence (Eq. 31) a much more adequate measure of electrophysiological connectivity, because it removes the confounding effect of instantaneous dependence due to volume conduction and low spatial resolution (Pascual-Marqui 2007a).

Note that the measures of linear dependence defined by Eq. 15, Eq. 17, and Eq. 18 each have the form of a ratio of variances, which compares the residuals of different models (i.e. different dependent and independent variables). Under the assumption that the time series are wide-sense stationary, large sample distribution theory can be used to test the null hypothesis that a given measure of linear dependence is zero. Following the same methodology as in Geweke (1982), the asymptotic distributions are:

Eq. 32
$$\begin{cases} \text{Under } H_o: F_{X,Y}(\omega) = 0 \ , \ N_T \hat{F}_{X,Y}(\omega) \overset{a}{\sim} \chi^2(2pq) \\ \text{Under } H_o: F_{X \rightleftarrows Y}(\omega) = 0 \ , \ N_T \hat{F}_{X \rightleftarrows Y}(\omega) \overset{a}{\sim} \chi^2(pq) \\ \text{Under } H_o: F_{X \cdot Y}(\omega) = 0 \ , \ N_T \hat{F}_{X \cdot Y}(\omega) \overset{a}{\sim} \chi^2(pq) \end{cases}$$

## 4. Measures of linear dependence (coherence-type) between groups of multivariate time series

Consider the case of three multivariate time series $\mathbf{X}_{jt} \in \mathbb{R}^{p \times 1}$, $\mathbf{Y}_{jt} \in \mathbb{R}^{q \times 1}$, and $\mathbf{Z}_{jt} \in \mathbb{R}^{r \times 1}$, for discrete time $t = 0...N_T - 1$, with $j = 1...N_R$ denoting the *j*-th time segment.

The measures of linear dependence between the three multivariate time series are related in the usual way:

**Eq. 33** $\quad F_{X,Y,Z}(\omega) = F_{X \rightleftarrows Y \rightleftarrows Z}(\omega) + F_{X \cdot Y \cdot Z}(\omega)$

and are given by:

**Eq. 34** $\quad F_{X,Y,Z}(\omega) = \ln \dfrac{\left| \begin{pmatrix} S_{YY\omega} & 0 & 0 \\ 0 & S_{XX\omega} & 0 \\ 0 & 0 & S_{ZZ\omega} \end{pmatrix} \right|}{\left| \begin{pmatrix} S_{YY\omega} & S_{YX\omega} & S_{YZ\omega} \\ S_{XY\omega} & S_{XX\omega} & S_{XZ\omega} \\ S_{ZY\omega} & S_{ZX\omega} & S_{ZZ\omega} \end{pmatrix} \right|}$

**Eq. 35** $\quad F_{X \cdot Y \cdot Z}(\omega) = \ln \dfrac{\left| \mathrm{Re} \begin{pmatrix} S_{YY\omega} & 0 & 0 \\ 0 & S_{XX\omega} & 0 \\ 0 & 0 & S_{ZZ\omega} \end{pmatrix} \right|}{\left| \mathrm{Re} \begin{pmatrix} S_{YY\omega} & S_{YX\omega} & S_{YZ\omega} \\ S_{XY\omega} & S_{XX\omega} & S_{XZ\omega} \\ S_{ZY\omega} & S_{ZX\omega} & S_{ZZ\omega} \end{pmatrix} \right|}$

and:

**Eq. 36** $\quad F_{X \rightleftarrows Y \rightleftarrows Z}(\omega) = \ln \dfrac{\left| \mathrm{Re} \begin{pmatrix} S_{YY\omega} & S_{YX\omega} & S_{YZ\omega} \\ S_{XY\omega} & S_{XX\omega} & S_{XZ\omega} \\ S_{ZY\omega} & S_{ZX\omega} & S_{ZZ\omega} \end{pmatrix} \right| \Big/ \left| \mathrm{Re} \begin{pmatrix} S_{YY\omega} & 0 & 0 \\ 0 & S_{XX\omega} & 0 \\ 0 & 0 & S_{ZZ\omega} \end{pmatrix} \right|}{\left\{ \left| \begin{pmatrix} S_{YY\omega} & S_{YX\omega} & S_{YZ\omega} \\ S_{XY\omega} & S_{XX\omega} & S_{XZ\omega} \\ S_{ZY\omega} & S_{ZX\omega} & S_{ZZ\omega} \end{pmatrix} \right| \Big/ \left| \begin{pmatrix} S_{YY\omega} & 0 & 0 \\ 0 & S_{XX\omega} & 0 \\ 0 & 0 & S_{ZZ\omega} \end{pmatrix} \right| \right\}}$

Coherences for each type of measure of linear dependence in Eq. 33 are defined by the general relation (see e.g. Pierce 1982):

**Eq. 37** $\quad \rho^2(\omega) = 1 - \exp\left[ -F(\omega) \right]$

As previously argued, under the assumption that the time series are wide-sense stationary, large sample distribution theory can be used to test the null hypothesis that a given measure of linear dependence is zero. In this case, the asymptotic distributions are:

**Eq. 38** $\quad \begin{cases} Under \ H_o : F_{X,Y,Z}(\omega) = 0 \ , \ N_T \hat{F}_{X,Y,Z}(\omega) \overset{a}{\sim} \chi^2(2pq + 2pr + 2qr) \\ Under \ H_o : F_{X \rightleftarrows Y \rightleftarrows Z}(\omega) = 0 \ , \ N_T \hat{F}_{X \rightleftarrows Y \rightleftarrows Z}(\omega) \overset{a}{\sim} \chi^2(pq + pr + qr) \\ Under \ H_o : F_{X \cdot Y \cdot Z}(\omega) = 0 \ , \ N_T \hat{F}_{X \cdot Y \cdot Z}(\omega) \overset{a}{\sim} \chi^2(pq + pr + qr) \end{cases}$

The generalization of these definitions to any number of multivariate time series is straightforward.

It is important to emphasize here that these measures of linear dependence for groups of multivariate time series can be applied in the field of neurophysiology. In this case, the time series consist of electric neuronal activity at several brain locations, and the measures of dependence are interpreted as "connectivity" between locations. When considering several brain locations, these new measures can be used to test for the existence of distributed cortical networks, whose activity can be estimated with exact low resolution brain electromagnetic tomography (Pascual-Marqui 2007b).

## 5. Measures of linear dependence (coherence-type) between all univariate time series

A particular case of interest consists of measuring the linear dependence between all the univariate time series that form part of the vector time series. For instance, consider the vector time series $\mathbf{X}_{jt} \in \mathbb{R}^{p \times 1}$. Then the measures of linear dependence between all "$p$" univariate time series of $\mathbf{X}$ are:

**Eq. 39** $\quad F_{\mathbf{X},\mathbf{X}}(\omega) = F_{\mathbf{X} \rightleftarrows \mathbf{X}}(\omega) + F_{\mathbf{X} \cdot \mathbf{X}}(\omega)$

**Eq. 40** $\quad F_{\mathbf{X},\mathbf{X}}(\omega) = \ln \frac{|Diag(\mathbf{S}_{\mathbf{XX}\omega})|}{|\mathbf{S}_{\mathbf{XX}\omega}|}$

**Eq. 41** $\quad F_{\mathbf{X} \cdot \mathbf{X}}(\omega) = \ln \frac{|Diag(\mathbf{S}_{\mathbf{XX}\omega})|}{|Re(\mathbf{S}_{\mathbf{XX}\omega})|}$

**Eq. 42** $\quad F_{\mathbf{X} \rightleftarrows \mathbf{X}}(\omega) = F_{\mathbf{X},\mathbf{X}}(\omega) - F_{\mathbf{X} \cdot \mathbf{X}}(\omega) = \ln \frac{|Re(\mathbf{S}_{\mathbf{XX}\omega})|}{|\mathbf{S}_{\mathbf{XX}\omega}|}$

Coherences for each type of measure of linear dependence in Eq. 39 are defined by the general relation (see e.g. Pierce 1982):

**Eq. 43** $\quad \rho^2(\omega) = 1 - \exp[-F(\omega)]$

In Eq. 40 and Eq. 41, the notation $Diag(\mathbf{M})$ denotes a diagonal matrix formed by the diagonal elements of $\mathbf{M}$. Note that for Hermitian matrices, such as $\mathbf{S}_{\mathbf{XX}\omega}$, the diagonal elements are pure real, which implies that:

**Eq. 44** $\quad Diag(\mathbf{S}_{\mathbf{XX}\omega}) = Diag(Re(\mathbf{S}_{\mathbf{XX}\omega})) = Re(Diag(\mathbf{S}_{\mathbf{XX}\omega}))$

As a consistency check, it can easily be verified that when these definitions are applied to a vector time series with 2 components, the same results are obtained as in the case of two univariate time series (Eq. 21, Eq. 23, and Eq. 27).

Under the assumption that the time series are wide-sense stationary, large sample distribution theory can be used to test the null hypothesis that a given measure of linear dependence is zero. In this case, the asymptotic distributions are:

**Eq. 45** $\quad \begin{cases} Under\ H_o: F_{\mathbf{X},\mathbf{X}}(\omega) = 0\ ,\ N_T \hat{F}_{\mathbf{X},\mathbf{X}}(\omega) \stackrel{a}{\sim} \chi^2(p(p-1)) \\ Under\ H_o: F_{\mathbf{X} \rightleftarrows \mathbf{X}}(\omega) = 0\ ,\ N_T \hat{F}_{\mathbf{X} \rightleftarrows \mathbf{X}}(\omega) \stackrel{a}{\sim} \chi^2(p(p-1)/2) \\ Under\ H_o: F_{\mathbf{X} \cdot \mathbf{X}}(\omega) = 0\ ,\ N_T \hat{F}_{\mathbf{X} \cdot \mathbf{X}}(\omega) \stackrel{a}{\sim} \chi^2(p(p-1)/2) \end{cases}$

As a further consistency check, note that the test $H_o : F_{\mathbf{X} \cdot \mathbf{X}}(\omega) = 0$ corresponds to the classical case of testing if a real-valued correlation matrix is the identity matrix. The statistic given above is precisely the log-likelihood ratio statistic, which is asymptotically chi-square with the specified degrees of freedom (Kullback 1967).

## 6. Measures of nonlinear dependence (phase synchronization type) between two multivariate time series

The term "phase synchronization" has a very rigorous physics definition (see e.g. Rosenblum et al 1996). The basic idea behind this definition has been adapted and used to great advantage in the neurosciences (Tass et al 1998, Quian-Quiroga et al 2002, Pereda et al 2005, Stam et al 2007), as in for example, the analysis of pairs of time series of measured scalp electric potentials differences (i.e. EEG: electroencephalogram). Other equivalent descriptive names for "phase synchronization" that appear in the neurosciences are phase locking, phase locking value, phase locking index, phase coherence, and so on.

An informal definition for the statistical "phase synchronization" model will now be given. In order to simplify this informal definition even further, it will be assumed that there are two univariate stationary time series (i.e. $p = q = 1$) of interest. At a given discrete frequency $\omega$, the sample data in the frequency domain (using the discrete Fourier transform) is denoted as $x_{j\omega}, y_{j\omega} \in \mathbb{C}$, with $j = 1 \ldots N_R$ denoting the $j$-th time segment. If the phase difference $\Delta \varphi_j = \varphi_j^x - \varphi_j^y$ is "stable" over time segments $j$, regardless of the amplitudes, then there is a "connection" between the locations at which the measurements were made. A measure of stability of phase difference is precisely "phase synchronization". It can as well be defined for the non-stationary case, using concepts of time-varying instantaneous phase, and defining stability over time (instead of stability over time segments).

In the case of univariate time series, i.e. $p = q = 1$, phase synchronization can be viewed as the modulus (absolute value) of the complex valued (Hermitian) coherency between the _normalized_ Fourier transforms. These variables are normalized prior to the coherency calculation in order to remove from the outset any amplitude effect, leaving only phase information. This normalization operation is highly nonlinear.

The modulus of the coherency is used as a measure for phase synchronization because it is conveniently bounded in the range zero (no synchronization) to one (perfect synchronization).

Based on the foregoing arguments, a natural definition for the measures of nonlinear dependence (phase synchronization type) between two multivariate time series is exactly the same definitions as developed in the previous sections of this study, but applied to the phase-information cross-spectra (Eq. 7 to Eq. 12). The phase-information cross-spectra are based on normalized Fourier transform vectors, which is the particular requirement in this case (without amplitude information).

For two multivariate time series, the measure of nonlinear dependence $G_{X,Y}(\omega)$ is expressed as the sum of lagged nonlinear dependence $G_{X \rightleftarrows Y}(\omega)$ and instantaneous nonlinear dependence $G_{X \cdot Y}(\omega)$:

**Eq. 46** $\quad G_{X,Y}(\omega) = G_{X \rightleftarrows Y}(\omega) + G_{X \cdot Y}(\omega)$

with:

**Eq. 47** $\quad G_{X,Y}(\omega) = \ln \dfrac{\left| \begin{pmatrix} S_{\bar{Y}\bar{Y}\omega} & \mathbf{0} \\ \mathbf{0}^T & S_{\bar{X}\bar{X}\omega} \end{pmatrix} \right|}{\left| \begin{pmatrix} S_{\bar{Y}\bar{Y}\omega} & S_{\bar{Y}\bar{X}\omega} \\ S_{\bar{X}\bar{Y}\omega} & S_{\bar{X}\bar{X}\omega} \end{pmatrix} \right|}$

**Eq. 48** $\quad G_{X \cdot Y}(\omega) = \ln \dfrac{\left| \mathrm{Re} \begin{pmatrix} S_{\bar{Y}\bar{Y}\omega} & \mathbf{0} \\ \mathbf{0}^T & S_{\bar{X}\bar{X}\omega} \end{pmatrix} \right|}{\left| \mathrm{Re} \begin{pmatrix} S_{\bar{Y}\bar{Y}\omega} & S_{\bar{Y}\bar{X}\omega} \\ S_{\bar{X}\bar{Y}\omega} & S_{\bar{X}\bar{X}\omega} \end{pmatrix} \right|}$

and:

**Eq. 49** $\quad G_{X \rightleftarrows Y}(\omega) = G_{X,Y}(\omega) - G_{X \cdot Y}(\omega) = \ln \dfrac{\left\{ \left| \mathrm{Re} \begin{pmatrix} S_{\bar{Y}\bar{Y}\omega} & S_{\bar{Y}\bar{X}\omega} \\ S_{\bar{X}\bar{Y}\omega} & S_{\bar{X}\bar{X}\omega} \end{pmatrix} \right| \Big/ \left| \mathrm{Re} \begin{pmatrix} S_{\bar{Y}\bar{Y}\omega} & \mathbf{0} \\ \mathbf{0}^T & S_{\bar{X}\bar{X}\omega} \end{pmatrix} \right| \right\}}{\left\{ \left| \begin{pmatrix} S_{\bar{Y}\bar{Y}\omega} & S_{\bar{Y}\bar{X}\omega} \\ S_{\bar{X}\bar{Y}\omega} & S_{\bar{X}\bar{X}\omega} \end{pmatrix} \right| \Big/ \left| \begin{pmatrix} S_{\bar{Y}\bar{Y}\omega} & \mathbf{0} \\ \mathbf{0}^T & S_{\bar{X}\bar{X}\omega} \end{pmatrix} \right| \right\}}$

In Eq. 47, Eq. 48, and Eq. 49, the Hermitian covariance matrices are defined for the normalized discrete Fourier transform vectors (Eq. 7 to Eq. 12).

All three measures are non-negative. They take the value zero only when there is independence of the pertinent type (lagged, instantaneous, or both).

These measures of nonlinear dependence can be associated with measures phase synchronization $\varphi$ as follows.

The phase synchronization between two multivariate time series is:

**Eq. 50** $\quad \varphi_{X,Y}^2(\omega) = 1 - \exp(-G_{X,Y}(\omega)) = 1 - \dfrac{\left| \begin{pmatrix} S_{\bar{Y}\bar{Y}\omega} & S_{\bar{Y}\bar{X}\omega} \\ S_{\bar{X}\bar{Y}\omega} & S_{\bar{X}\bar{X}\omega} \end{pmatrix} \right|}{\left| \begin{pmatrix} S_{\bar{Y}\bar{Y}\omega} & \mathbf{0} \\ \mathbf{0}^T & S_{\bar{X}\bar{X}\omega} \end{pmatrix} \right|}$

The instantaneous phase synchronization between two multivariate time series is:

**Eq. 51** $\quad \varphi_{X \cdot Y}^2(\omega) = 1 - \exp(-G_{X \cdot Y}(\omega)) = 1 - \dfrac{\left| \mathrm{Re} \begin{pmatrix} S_{\bar{Y}\bar{Y}\omega} & S_{\bar{Y}\bar{X}\omega} \\ S_{\bar{X}\bar{Y}\omega} & S_{\bar{X}\bar{X}\omega} \end{pmatrix} \right|}{\left| \mathrm{Re} \begin{pmatrix} S_{\bar{Y}\bar{Y}\omega} & \mathbf{0} \\ \mathbf{0}^T & S_{\bar{X}\bar{X}\omega} \end{pmatrix} \right|}$

The lagged phase synchronization between two multivariate time series is:

Eq. 52
$$\varphi^2_{X\rightleftarrows Y}(\omega) = 1 - \exp(-G_{X\rightleftarrows Y}(\omega)) = 1 - \frac{\left\{ \left| \begin{pmatrix} S_{\breve{Y}\breve{Y}\omega} & S_{\breve{Y}\breve{X}\omega} \\ S_{\breve{X}\breve{Y}\omega} & S_{\breve{X}\breve{X}\omega} \end{pmatrix} \right| \middle/ \left| \begin{pmatrix} S_{\breve{Y}\breve{Y}\omega} & \mathbf{o} \\ \mathbf{o}^T & S_{\breve{X}\breve{X}\omega} \end{pmatrix} \right| \right\}}{\left\{ \left| \mathrm{Re} \begin{pmatrix} S_{\breve{Y}\breve{Y}\omega} & S_{\breve{Y}\breve{X}\omega} \\ S_{\breve{X}\breve{Y}\omega} & S_{\breve{X}\breve{X}\omega} \end{pmatrix} \right| \middle/ \left| \mathrm{Re} \begin{pmatrix} S_{\breve{Y}\breve{Y}\omega} & \mathbf{o} \\ \mathbf{o}^T & S_{\breve{X}\breve{X}\omega} \end{pmatrix} \right| \right\}}$$

The phase synchronization between two multivariate time series $\varphi^2_{X,Y}(\omega)$ given by Eq. 50 corresponds to the square of the "general phase synchronization" previously defined in Pascual-Marqui (2007a; see Eq. 15 therein).

In order to illustrate and further motivate these measures of nonlinear dependence, a detailed analysis for the simple case of two univariate time series is presented.

In the case that the two time series are univariate, the measure of nonlinear dependence $G_{X,Y}(\omega)$ in Eq. 47 is:

Eq. 53
$$G_{X,Y}(\omega) = \ln \frac{1}{1 - \left[\mathrm{Re}(s_{\breve{x}\breve{y}\omega})\right]^2 - \left[\mathrm{Im}(s_{\breve{x}\breve{y}\omega})\right]^2} = -\ln(1 - \varphi^2_{X,Y}(\omega))$$

with phase synchronization:

Eq. 54
$$\varphi^2_{X,Y}(\omega) = \left[\mathrm{Re}(s_{\breve{x}\breve{y}\omega})\right]^2 + \left[\mathrm{Im}(s_{\breve{x}\breve{y}\omega})\right]^2 = \left| \frac{1}{N_R} \sum_{j=1}^{N_R} \breve{x}_{j\omega} \breve{y}^*_{j\omega} \right|^2$$

Note that by definition, due to the normalization, $s_{\breve{x}\breve{x}\omega} = s_{\breve{y}\breve{y}\omega} = 1$. In Eq. 54, $\varphi_{X,Y}$ is the classical measure of phase synchronization.

The measure of instantaneous nonlinear dependence is:

Eq. 55
$$G_{X\cdot Y}(\omega) = \ln \frac{1}{1 - \left[\mathrm{Re}(s_{\breve{x}\breve{y}\omega})\right]^2} = -\ln(1 - \varphi^2_{X\cdot Y}(\omega))$$

with instantaneous phase synchronization:

Eq. 56
$$\varphi^2_{X\cdot Y}(\omega) = \left[\mathrm{Re}(s_{\breve{x}\breve{y}\omega})\right]^2$$

which, not surprisingly, is directly related to the real part of the complex valued coherency of the normalized time series.

Finally, in the particular case of univariate time series, the measure of lagged nonlinear dependence is:

Eq. 57
$$G_{X\rightleftarrows Y}(\omega) = \ln \frac{1 - \left[\mathrm{Re}(s_{\breve{x}\breve{y}\omega})\right]^2}{1 - \left[\mathrm{Re}(s_{\breve{x}\breve{y}\omega})\right]^2 - \left[\mathrm{Im}(s_{\breve{x}\breve{y}\omega})\right]^2} = -\ln(1 - \varphi^2_{X\rightleftarrows Y}(\omega))$$

with lagged phase synchronization:

Eq. 58
$$\varphi^2_{X\rightleftarrows Y}(\omega) = \frac{\left[\mathrm{Im}(s_{\breve{x}\breve{y}\omega})\right]^2}{1 - \left[\mathrm{Re}(s_{\breve{x}\breve{y}\omega})\right]^2}$$

The lagged phase synchronization between two univariate time series $\varphi_{X\rightleftarrows Y}(\omega)$ given by Eq. 58 corresponds to the "general lagged phase synchronization" (i.e. the "zero-lag removed" general phase synchronization)" previously defined in Pascual-Marqui (2007a), see Eq. 33 therein.

It is worth stressing the asymmetry in the results for the instantaneous phase synchronization $\varphi^2_{X \cdot Y}(\omega)$ (Eq. 56) and the lagged phase synchronization $\varphi^2_{X \rightleftarrows Y}(\omega)$ (Eq. 58). While the instantaneous phase synchronization is the real part of the complex valued coherency for the normalized time series, the lagged phase synchronization *is not* the imaginary part. Ideally, the lagged phase synchronization is a measure that is less affected by instantaneous nonlinear dependence.

In our previous related study (Pascual-Marqui 2007a), the definition given there for the "zero-lag removed general phase synchronization" (see Eq. 28 therein) was:

Eq. 59
$$PS_{GL} = \rho_{GL} = \sqrt{1 - \frac{\left|\begin{pmatrix} \mathbf{S}_{\bar{Y}\bar{Y}\omega} & \mathbf{S}_{\bar{Y}\bar{X}\omega} \\ \mathbf{S}_{\bar{X}\bar{Y}\omega} & \mathbf{S}_{\bar{X}\bar{X}\omega} \end{pmatrix}\right|}{\left|\mathrm{Re}\begin{pmatrix} \mathbf{S}_{\bar{Y}\bar{Y}\omega} & \mathbf{S}_{\bar{Y}\bar{X}\omega} \\ \mathbf{S}_{\bar{X}\bar{Y}\omega} & \mathbf{S}_{\bar{X}\bar{X}\omega} \end{pmatrix}\right|}}$$

The new definition given here for the lagged phase synchronization $\varphi^2_{X \rightleftarrows Y}(\omega)$ is given by Eq. 52. Both definitions (Eq. 52 and Eq. 59) are identical for the case of two univariate time series. However, they are different for the multivariate case. Whereas the old definition in Eq. 59 lumps together all variables from X and Y, the new definition given here in Eq. 52 conserves the multivariate structure of the two multivariate time series. The improvement of the new lagged phase synchronization in Eq. 52 is that it measures the lagged nonlinear dependence between the two multivariate time series without being affected by the covariance structure within each multivariate time series. The shortcoming of the old definition from our previous study (Pascual-Marqui 2007a), shown in Eq. 59, is that it is contaminated by the dependence structures of the univariate time series within **X** and within **Y**.

## 7. *Measures of nonlinear dependence (phase synchronization type) between groups of multivariate time series*

Consider the case of three multivariate time series $\mathbf{X}_{jt} \in \mathbb{R}^{p \times 1}$, $\mathbf{Y}_{jt} \in \mathbb{R}^{q \times 1}$, and $\mathbf{Z}_{jt} \in \mathbb{R}^{r \times 1}$, for discrete time $t = 0 \ldots N_T - 1$, with $j = 1 \ldots N_R$ denoting the *j*-th time segment.

The measures of nonlinear dependence between the three multivariate time series are related in the usual way:

Eq. 60 $\qquad G_{X,Y,Z}(\omega) = G_{X \rightleftarrows Y \rightleftarrows Z}(\omega) + G_{X \cdot Y \cdot Z}(\omega)$

and are given by:

**Eq. 61**
$$G_{X,Y,Z}(\omega) = \ln \frac{\left| \begin{pmatrix} S_{\bar{Y}\bar{Y}\omega} & 0 & 0 \\ 0 & S_{\bar{X}\bar{X}\omega} & 0 \\ 0 & 0 & S_{\bar{Z}\bar{Z}\omega} \end{pmatrix} \right|}{\left| \begin{pmatrix} S_{\bar{Y}\bar{Y}\omega} & S_{\bar{Y}\bar{X}\omega} & S_{\bar{Y}\bar{Z}\omega} \\ S_{\bar{X}\bar{Y}\omega} & S_{\bar{X}\bar{X}\omega} & S_{\bar{X}\bar{Z}\omega} \\ S_{\bar{Z}\bar{Y}\omega} & S_{\bar{Z}\bar{X}\omega} & S_{\bar{Z}\bar{Z}\omega} \end{pmatrix} \right|}$$

**Eq. 62**
$$G_{X.Y.Z}(\omega) = \ln \frac{\left| \mathrm{Re} \begin{pmatrix} S_{\bar{Y}\bar{Y}\omega} & 0 & 0 \\ 0 & S_{\bar{X}\bar{X}\omega} & 0 \\ 0 & 0 & S_{\bar{Z}\bar{Z}\omega} \end{pmatrix} \right|}{\left| \mathrm{Re} \begin{pmatrix} S_{\bar{Y}\bar{Y}\omega} & S_{\bar{Y}\bar{X}\omega} & S_{\bar{Y}\bar{Z}\omega} \\ S_{\bar{X}\bar{Y}\omega} & S_{\bar{X}\bar{X}\omega} & S_{\bar{X}\bar{Z}\omega} \\ S_{\bar{Z}\bar{Y}\omega} & S_{\bar{Z}\bar{X}\omega} & S_{\bar{Z}\bar{Z}\omega} \end{pmatrix} \right|}$$

**Eq. 63**
$$G_{X \rightleftarrows Y \rightleftarrows Z}(\omega) = \ln \frac{\left| \mathrm{Re} \begin{pmatrix} S_{\bar{Y}\bar{Y}\omega} & 0 & 0 \\ 0 & S_{\bar{X}\bar{X}\omega} & 0 \\ 0 & 0 & S_{\bar{Z}\bar{Z}\omega} \end{pmatrix} \right| / \left| \mathrm{Re} \begin{pmatrix} S_{\bar{Y}\bar{Y}\omega} & 0 & 0 \\ 0 & S_{\bar{X}\bar{X}\omega} & 0 \\ 0 & 0 & S_{\bar{Z}\bar{Z}\omega} \end{pmatrix} \right|}{\left\{ \left| \begin{pmatrix} S_{\bar{Y}\bar{Y}\omega} & S_{\bar{Y}\bar{X}\omega} & S_{\bar{Y}\bar{Z}\omega} \\ S_{\bar{X}\bar{Y}\omega} & S_{\bar{X}\bar{X}\omega} & S_{\bar{X}\bar{Z}\omega} \\ S_{\bar{Z}\bar{Y}\omega} & S_{\bar{Z}\bar{X}\omega} & S_{\bar{Z}\bar{Z}\omega} \end{pmatrix} \right| / \left| \begin{pmatrix} S_{\bar{Y}\bar{Y}\omega} & 0 & 0 \\ 0 & S_{\bar{X}\bar{X}\omega} & 0 \\ 0 & 0 & S_{\bar{Z}\bar{Z}\omega} \end{pmatrix} \right| \right\}}$$

Phase synchronization for each type of measure of linear dependence in Eq. 60 can be defined by the general relation (see e.g. Pierce 1982):

**Eq. 64** $\quad \varphi^2(\omega) = 1 - \exp\left[-G(\omega)\right]$

The generalization of these definitions to any number of multivariate time series is straightforward.

It is important to emphasize here that these measures of nonlinear dependence for groups of multivariate time series can be applied in the field of neurophysiology. In this case, the time series consist of electric neuronal activity at several brain locations, and the measures of dependence are interpreted as "connectivity" between locations. When considering several brain locations, these new measures can be used to test for the existence of distributed cortical networks, whose activity can be estimated with exact low resolution brain electromagnetic tomography (Pascual-Marqui 2007b).

## 8. *Measures of nonlinear dependence (phase synchronization type) between all univariate time series*

A particular case of interest consists of measuring the nonlinear dependence between all the univariate time series that form part of the vector time series. For instance, consider the vector time series $\mathbf{X}_{jt} \in \mathbb{R}^{p \times 1}$. In this case, since each univariate time series on its own is of interest, each one must be normalized. For this particular purpose we adopt the definition:

Eq. 65 $$\widehat{\mathbf{X}}_{j\omega} = \left[ Diag\left( \mathbf{X}_{j\omega} \mathbf{X}_{j\omega}^* \right) \right]^{-1/2} \mathbf{X}_{j\omega}$$

which normalizes each variable. The corresponding covariance matrix is:

Eq. 66 $$\mathbf{S}_{\widehat{X}\widehat{X}\omega} = \frac{1}{N_R} \sum_{j=1}^{N_R} \widehat{\mathbf{X}}_{j\omega} \widehat{\mathbf{X}}_{j\omega}^*$$

Then the measures of nonlinear dependence between all "$p$" univariate time series of **X** are:

Eq. 67 $$G_{X,X}(\omega) = G_{X \rightleftarrows X}(\omega) + G_{X \cdot X}(\omega)$$

Eq. 68 $$G_{X,X}(\omega) = -\ln \left| \mathbf{S}_{\widehat{X}\widehat{X}\omega} \right|$$

Eq. 69 $$G_{X \cdot X}(\omega) = -\ln \left| \operatorname{Re}\left( \mathbf{S}_{\widehat{X}\widehat{X}\omega} \right) \right|$$

Eq. 70 $$G_{X \rightleftarrows X}(\omega) = G_{X,X}(\omega) - G_{X \cdot X}(\omega) = \ln \frac{\left| \operatorname{Re}\left( \mathbf{S}_{\widehat{X}\widehat{X}\omega} \right) \right|}{\left| \mathbf{S}_{\widehat{X}\widehat{X}\omega} \right|}$$

Phase synchronization for each type of measure of linear dependence in Eq. 67 can be defined by the general relation (see e.g. Pierce 1982):

Eq. 71 $$\varphi^2(\omega) = 1 - \exp\left[ -G(\omega) \right]$$

As a consistency check, it can easily be verified that when these definitions are applied to a vector time series with 2 components, the same results are obtained as in the case of two univariate time series (Eq. 53, Eq. 55, and Eq. 57).

## 9. Conclusions

1. Previous related work (Pascual-Marqui 2007a) was limited to measures of dependence between two multivariate time series. This study generalizes the definitions to include measures of dependence between any number of multivariate time series.

2. Previous measures for lagged dependence between two vector time series (Pascual-Marqui 2007a) were inadequately affected by the dependence structure of the univariate time series within each vector time series. This study adequately partials out the dependence structures within each vector time series.

3. A new measure for instantaneous linear and non-linear dependence is introduced.

4. The measures of dependence introduced here have been developed for discrete frequency components. However, they can as well be applied to any frequency band, defined as a set of discrete frequencies (which can even be disjoint). In this case, the Hermitian covariance matrices to be used in the equations for the measures of dependence should now correspond to the pooled matrices (i.e. the average Hermitian covariance over all discrete frequencies in the set defining the frequency band).

5. Inference methods for the measures of linear dependence are described.

6. All the measures of dependence can be based on any form of time-varying Fourier transforms or wavelets, such as, for instance, Gabor or Morlet transforms.

7. The new measures of dependence between any number of multivariate time series can be applied to the study of brain electrical activity, which can be estimated non-invasively from EEG/MEG recordings with methods such as eLORETA (Pascual-Marqui 2007b). When considering several brain locations jointly, these new measures can be used to test for the existence of distributed cortical networks. Previous methodology explores the connections between all possible pairs of locations, while the new "network approach" can test the joint dependence of several locations.

## *Appendix 1: Zero-lag contribution to coherence and phase synchronization: problem description*

In some fields of application, the coherence or phase synchronization between two time series corresponding to two different spatial locations is interpreted as a measure of the "connectivity" between those two locations.

For example, consider the time series of scalp electric potential differences (EEG: electroencephalogram) at two locations. The coherence or phase synchronization is interpreted by some researchers as a measure of "connectivity" between the underlying cortices (see e.g. Nolte et al 2004 and Stam et al 2007).

However, even if the underlying cortices are not actually connected, significantly high coherence or phase synchronization might still occur due to the volume conduction effect: activity at any cortical area will be observed instantaneously (zero-lag) by all scalp electrodes.

As a possible solution to this problem, the electric neuronal activity distributed throughout the cortex can be estimated from the EEG by using imaging techniques such as standardized or exact low resolution brain electromagnetic tomography (sLORETA, eLORETA) (Pascual-Marqui et al 2002; Pascual-Marqui 2007b). At each voxel in the cortical grey matter, a 3-component vector time series is computed, corresponding to the current density vector with dipole moments along axes *X*, *Y*, and *Z*. This tomography has the unique properties of being linear, of having zero localization error, but of having low spatial resolution. Due to such spatial "blurring", the time series will again suffer from non-physiological inflated values of zero-lag coherence and phase synchronization.

Formally, consider two different spatial locations where there is no actual activity. However, due to a third truly active location, and because of low spatial resolution (or volume conductor type effect), there is some measured activity at these locations:

**Eq. 72:** $\begin{cases} \mathbf{X}_{jt} = \mathbf{C}\mathbf{Z}_{jt} + \boldsymbol{\varepsilon}_{jt}^{x} \\ \mathbf{Y}_{jt} = \mathbf{D}\mathbf{Z}_{jt} + \boldsymbol{\varepsilon}_{jt}^{y} \end{cases}$

where $\mathbf{Z}_{jt}$ is the time series of the truly active location; $\mathbf{C}$ and $\mathbf{D}$ are matrices determined by the properties of the low spatial resolution problem; and $\varepsilon_{jt}^{x}$ and $\varepsilon_{jt}^{y}$ are independent and identically distributed random white noise.

In this model, although $\mathbf{X}$ and $\mathbf{Y}$ are not "connected", coherence and phase synchronization will indicate some connection, due to zero-lag spatial blurring.

Things can get even worse due to the zero-lag effect. Suppose that two time series are measured under two different conditions in which the zero-lag blurring effect is constant. The goal is to perform a statistical test to compare if there is a change in connectivity. Since the zero-lag effect is the same in both conditions, then it should seemingly not account for any significant difference in coherence or phase synchronization. However, this might be very misleading. In the model in Eq. 72, a simple increase in the signal to noise ratio (e.g. by increasing the norms of $\mathbf{C}$ and $\mathbf{D}$) will produce an increase in coherence and phase synchronization, due again to the zero-lag effect. This example shows that the zero-lag effect can render meaningless a comparison of two or more conditions.

## *Acknowledgements*